\documentclass[aps,pra,onecolumn,groupedaddress]{revtex4-1}
\usepackage{graphicx}
\usepackage{color}
\usepackage{dcolumn}
\usepackage{amsmath}
\usepackage{mathtools}
\usepackage{amssymb}
\usepackage{amsfonts}
\usepackage{bm}
\usepackage{epstopdf}
\usepackage{picture}
\usepackage{enumitem}
\usepackage{afterpage}
\usepackage{placeins}
\usepackage{float}
\usepackage{caption}
\usepackage{cases}
\usepackage{xr}

\setlist[enumerate]{label*=\arabic*.}

\newcommand{\avec}{\mathbf{a}}

\newcommand{\fvec}{\mathbf{f}}

\newcommand{\Kmat}{\mathbf{K}}

\newcommand{\Zmat}{\mathbf{Z}}
\newcommand{\Gmat}{\mathbf{G}}

\begin{document}

\title{Cold damping of levitated optically coupled nanoparticles}
\author{Vojt\v{e}ch Li\v{s}ka}
\author{Tereza Zem\'ankov\'a}
\author{Vojt\v{e}ch Svak}
\author{Petr J\'{a}kl}
\author{Jan Je\v{z}ek}
\author{Martin Br\'{a}neck\'y}
\author{Stephen H. Simpson}
\author{Pavel Zem\'anek}
\author{Oto Brzobohat\'y}
\email{otobrzo@isibrno.cz}

\affiliation{The Czech Academy of Sciences, Institute of Scientific Instruments, Kr\'{a}lovopolsk\'{a} 147, 612 64 Brno, Czech Republic}

\begin{abstract}
\noindent Methods for controlling the motion of single particles, optically levitated in vacuum, have developed rapidly in recent years.
The technique of \textit{cold damping} makes use of feedback-controlled, electrostatic forces to increase dissipation without introducing additional thermal fluctuations. 
This process has been instrumental in the ground-state cooling of individual electrically charged nanoparticles. 
Here we show that the same method can be applied to a pair of nanoparticles, coupled by optical binding forces. 
These optical binding forces are about three orders of magnitude stronger than typical Coulombic inter-particle force and result in a coupled motion of both nanoparticles characterized by a pair of normal modes. 
We demonstrate cold damping of these normal modes, either independently or simultaneously, to sub-Kelvin temperatures at pressures of $5\times10^{-3}$ mbar. 
Experimental observations are captured by a theoretical model which we use to survey the parameter space more widely and to quantify the limits imposed by measurement noise and time delays. Our work paves the way for the study of quantum interactions between meso-scale particles and the exploration of multiparticle entanglement in levitated optomechanical systems.
\end{abstract}

\maketitle

\section{Introduction}
Due to its isolation from the environment, a single nanoparticle, optically levitated in an ultrahigh vacuum, provides a promising experimental platform for weak force sensing \cite{Ranjit_PRA_2016, Hempston_APL_2017, Hebestreit_PRL_2018_2} and for testing fundamental physics at the boundary between the classical and quantum regimes \cite{Svak_NC_2018, Rondin_NatNano_2017}. 
With the recent achievement of ground state cooling for individual nanoparticles \cite{Delic_Sci_2020, Magrini_Nature_2021, Tebbenjohanns_Nature_2021}, levitational optomechanics opens up opportunities for experimentally exploring quantum effects at previously unattainable length scales and, potentially, for designing sensors with quantum-enhanced sensitivity. 
Extending levitational optomechanics to arrays of multiple, interacting particles is an exciting new development, still in its infancy \cite{DelicScience22}, which promises to open new research directions in quantum gravity\cite{Marletto_PRL_2017},  quantum friction measurements \cite{Zhao_PRL_2012}, dark matter detection \cite{Moore_QST_2021} or probing quantum correlations and entanglement \cite{Rudolph_PRL_2022}. 

To fully exploit the potential of these optomechanical arrays requires both understanding and control of the interaction forces acting between the particles. 
It also requires new protocols capable of cooling the multiple degrees of freedom defining these complex systems. The number of experimental demonstrations of optical trapping of interacting multiple particles is currently very limited. 
Bykov et al. reported long-range optical binding of multiple levitated microparticles mediated by intermodal scattering and interference in the evacuated core of a hollow-core photonic crystal fibre~\cite{Bykov_LSA_2018}, while Arita et al. demonstrated optical binding between two rotating chiral microparticles confined in vacuum in independent circularly polarised optical traps~\cite{Arita_Optica_18}. 

The tunable longitudinal \cite{Svak_Optica_21} and lateral \cite{DelicScience22} optical interactions between levitated nanoparticles have been demonstrated only very recently. 
An all-optical cold damping cooling approach \cite{SimmonsBIOPHYJ96,Kamba_OE_2022} has been very recently demonstrated for non-interacting nanoparticles \cite{Vijayan_NatNano_2022} while a sympathetic cooling scheme, inspired by experiments of cooling motion of atoms and ions \cite{Myatt_PRL_1997, Larson_PRL_1986}, has been applied for optically interacting rotating birefringent microparticles \cite{Arita_Optica_2022}.
Finally, the sympathetic cooling of two particles, coupled by a Coulombic interaction, and levitated in a Paul trap has been shown recently \cite{Bykov_Optica_23, Penny_PhysRevResearch_2023}. In this latter case, the mechanical oscillatory frequencies were typically several orders of magnitude smaller than in the case of optical levitation.

Here, we demonstrate cooling of the motion of two optically bound~\cite{Wei_SciRep16} nanoparticles levitated in two parallel optical traps in a vacuum to the sub-Kelvin level. 
This all-optical interaction\cite{BurnsSCI90,TatarkovaPRL02,SingerJOSAB03,DholakiaRMP10,CizmarJPB10, DemergisNL12,YanNC14,Li_ACSNano_2018, CizmarLPL11,BrzobohatyAPL11} is mediated by light scattering and, in our case, offers the ability to precisely control the interactions between particles \cite{DelicScience22, Wei_SciRep16}. 
For cooling the motion of the particles, we employ a cold-damping scheme \cite{Tebbenjohanns_PRL_2019, Magrini_Nature_2021}
\begin{figure*}[htb]
    \centering
    \includegraphics[width = 0.9\textwidth]{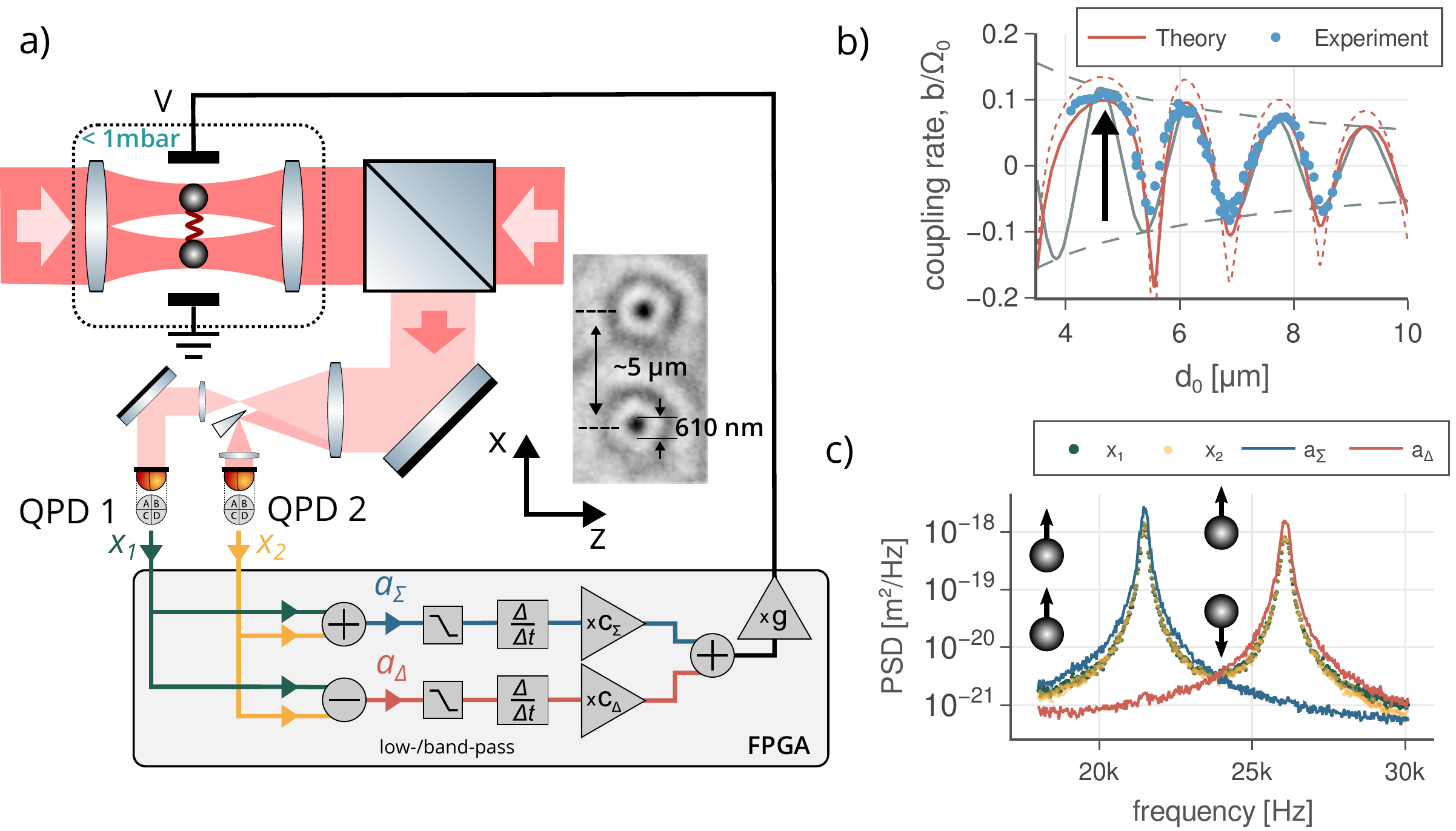}
    \caption{\label{fig1} Experimental geometry and observations. a) Two silica nanoparticles levitated in two parallel optical traps created by two pairs of interfering counter-propagating Gaussian beams. The position of each particle is monitored by the corresponding QPD. Independently, for QPD's calibration,  the $x-z$ plane was imaged on an ultra-fast CMOS camera. The box illustrates the feedback loop using FPGA processing. b)  The coupling rate can be controlled by varying the trap separation, $d_0$. The black arrow marks the separation between the traps used in our experiments. c) PSDs of the particle coordinates, $x_1$ and $x_2$, plotted together with PSDs of the center of mass $a_\Sigma$ and the breathing modes $a_\Delta$ for the coupled harmonic oscillators. }
\end{figure*}
which makes use of an external electric field to induce controllable, electrostatic forces on the pair of charged nanoparticles. This approach has previously been shown to be the most efficient feedback-based cooling method currently available  \cite{Penny_PRA_2021, Iwasaki_PRA_2019} and is, in theory, capable of preparing entangled states of interacting particles or acting as an ultra-sensitive sensor of the gradient of a probed external force field \cite{Rudolph_PRL_2022}.

\section{Experimental geometry}
In our experiments, two silica nanoparticles are levitated in two parallel optical traps created by two pairs of interfering counter-propagating Gaussian beams, see Fig. \ref{fig1}a). 
We employ a nebulizer to spray nanoparticles directly into the trapping region in the vacuum chamber. After successfully trapping two nanoparticles, we evacuated the vacuum chamber to sub-millibar pressure and monitored the motion of the particles using two custom-made quadrant photodiodes (QPDs, Hamamatsu G6849) with homodyne detection of the light transmitted through the optical traps, which has the particle motion encoded in its interference pattern, see Fig.~\ref{fig1}a).
In addition, we illuminated the particles along $y$ axis with a weak and wide laser beam ($\lambda = 532$\,nm,  beam waist $w_0 \approx 50\,\mu$m) and imaged the scattering patterns of both nanoparticles

with a microscope, onto a fast carefully calibrated CMOS camera (Vision Research Phantom V611),  see the colour inverted image of nanoparticles in Fig.~\ref{fig1}a). The frame rate was set to 200\,kHz and, typically, we recorded at least 100 000 frames, tracking the motion of the nanoparticles, to obtain sufficiently long trajectories for the analysis of their stochastic dynamics. 
This provided us with information about the in-plane $x$ and $z$ coordinates.
The record gave us quantitative, properly calibrated information about the amplitude of particles' motion without any prior knowledge of the particles' properties~\cite{Svak_NC_2018}. 
And by comparing parallel records from the camera and QPDs we calibrated the QPDs signals too~\cite{Svak_NC_2018}. 

The radius of levitated nanoparticles ($\rho = 305$\,nm) was significantly smaller than the wavelength of trapping beams and their beam waists ($\lambda = 1550$\,nm, $w_0 = 1.5\,\mu$m, $\rho/\lambda \approx 0.2$) and therefore the nanoparticles behave as induced dipoles.  
Due to the particular symmetry of our system and our use of relatively wide counter-propagating beams the optical binding force acts mainly along the vector connecting the particles ($x$ axis) which is perpendicular to the beam axes [i.e. $z$ axis in Fig. \ref{fig1}a)]. 
We note that our transversal direction of the binding force differs from that which acts prominently along the optical axis in parallel optical tweezers \cite{DelicScience22}.

The phase-coherent optical traps were generated by first-order diffraction from a digital micromirror device (DMD, Vialux). The total trapping power $2P = 140$ mW was split into two independent optical traps in a vacuum chamber, which allows us to independently set the trapping stiffness of both traps. The direction of polarization of the beams was controlled with a pair of half-wave plates and set along $y$ axis to maximize the scattering along $x$ axis and thus maximize the light-induced particle-particle interaction (optical binding). 
In the dipole approximation, the resulting optical binding forces oscillate periodically  \cite{Wei_SciRep16, BurnsPRL89} and decay
as  $F_{1,2} \propto \pm \sin(kd \pm \Delta\phi)/kd,$ where $k = 2\pi/\lambda$, $d$ is the inter-particle distance and $\Delta\phi$ denotes optical phase difference between the trapping laser beams incident on particle 1 and 2. Whenever $\Delta \phi \neq 0$, both binding forces have a component pointing in the same direction and forming a non-reciprocal and non-conservative interaction \cite{DelicScience22}.
Using the DMD we set $\Delta\phi = 0$ to ensure that the non-conservative part of the interaction \cite{DelicScience22} is negligible and set the distance between the optical traps to be approximately $d_0 = 5\,\mu$m to maximize the conservative part of the optical binding force, see Fig.~\ref{fig1}b).

Since the non-conservative part of the interaction is negligible, and the particles remain within the linear range of the traps, our system can be described as a pair of coupled oscillators  \cite{Svak_Optica_21} and thus their motion can be described in terms of a linear combination of two normal modes (Supplemental II A) i.e. the centre-of-mass (CoM) mode, with amplitude $a_{\Sigma} = (x_1 + x_2)/2$ and the breathing (BR) mode, with amplitude $a_\Delta = (x_1 - x_2)/2$. This can be clearly seen in the power spectral density of the particles motion, Fig.~\ref{fig1}c, which features two distinct resonant frequencies. 
The lower frequency resonance corresponds to the CoM mode, $\Omega_{\Sigma}/2\pi = (\Omega_0 - b)/2\pi \approx 21$\,kHz, and the higher corresponds to BR, $\Omega_{\Delta}/2\pi = (\Omega_0 +b)/2\pi \approx  26$\,kHz, where $\Omega_0$ is the resonant frequency of an individual trap. 
The parameter, $b$, is the coupling rate of the optical interaction, given by the derivative of the binding force evaluated at the stable positions of the particles (i.e. the derivative of $F_{1,2} \propto \cos(kx)/kx$ with respect to $x$, evaluated at $d$), Fig. \ref{fig1}b. 
We note that the separation of optical traps $d_0$ differs slightly from inter-particle separation $d$ due to the displacement of the particles in the traps, caused by the optical binding force.

Each particle is randomly charged. To estimate the magnitude of the Coulomb interaction between the particles we performed charge calibration\cite{Magrini_Nature_2021}. 
A sinusoidal voltage $V^\mathrm{d}(t) = V^\mathrm{d}_0 \sin \Omega_\mathrm{d} t$ with the driving frequency $\Omega_\mathrm{d}/2\pi = 50$\,kHz relatively close to the resonant frequencies of the normal modes ($\Omega_{\Sigma}$ and $\Omega_{\Delta}$). 
From the recorded trajectories of the particles, we calculated the power spectral densities for both normal modes and, by integrating them around driving frequency, we get variances  $\langle a_\Sigma^2 \rangle$ and $\langle a_\Delta^2 \rangle$. 
Assuming a model for a point charge in a parallel-plate capacitor, we get driving force acting on both modes in the form $F^\mathrm{d}_{\Sigma/\Delta}(t) = q_{\Sigma/\Delta}V^\mathrm{d}(t)/D,$ where $D = (198\pm 5)\,\mu$m is the distance between electrodes and   $q_\Sigma = (q_1 + q_2)/2$ and $q_\Delta = (q_1 - q_2)/2$ is the effective charge for the CoM and BR normal modes, respectively. These charges can be then expressed as 
\begin{equation}
    q_{\Sigma/\Delta} = \sqrt{2 \langle a_{\Sigma/\Delta}^2 \rangle} m D/V^\mathrm{d}_0 \sqrt{(\Omega_{\Sigma/\Delta}^2 - \Omega_\mathrm{d}^2)^2 + \xi_{\Sigma/\Delta}^2 \Omega_\mathrm{d}^2},    
\end{equation}
where $\xi_{\Sigma/\Delta}$ are the effective damping coefficients for the normal modes \cite{Svak_Optica_21}, $m$ is the mass of the particle and $V^\mathrm{d}_0 = 750$\,mV is amplitude of the applied voltage. 
The charges on the particles can then be written as $ q_{1,2} = (q_{\Sigma} \pm q_{\Delta})$, and the typical number of elementary charges on the particles was determined to be $N <  100$  and thus the magnitude of coupling rate $b/\Omega_0 = -q_1q_2/(8\pi\varepsilon_0 m \Omega_0^2d^3)< 10^{-4}$, for Coulomb interaction \cite{Rudolph_PRL_2022,DelicScience22} is  3 order of magnitude smaller than that one obtained for optical binding interaction, see Supplemental II C. 
This conclusion we confirmed experimentally when we suppressed the optical binding interaction by setting the polarization of the laser beams parallel to the $x$ axis and leaving only the Coulomb interaction. In this geometry, 
no mode splitting was detectable, showing that the Coulomb interaction is indeed negligible.

\begin{figure}[htb]
    \centering
    \includegraphics[width = 0.5\textwidth]{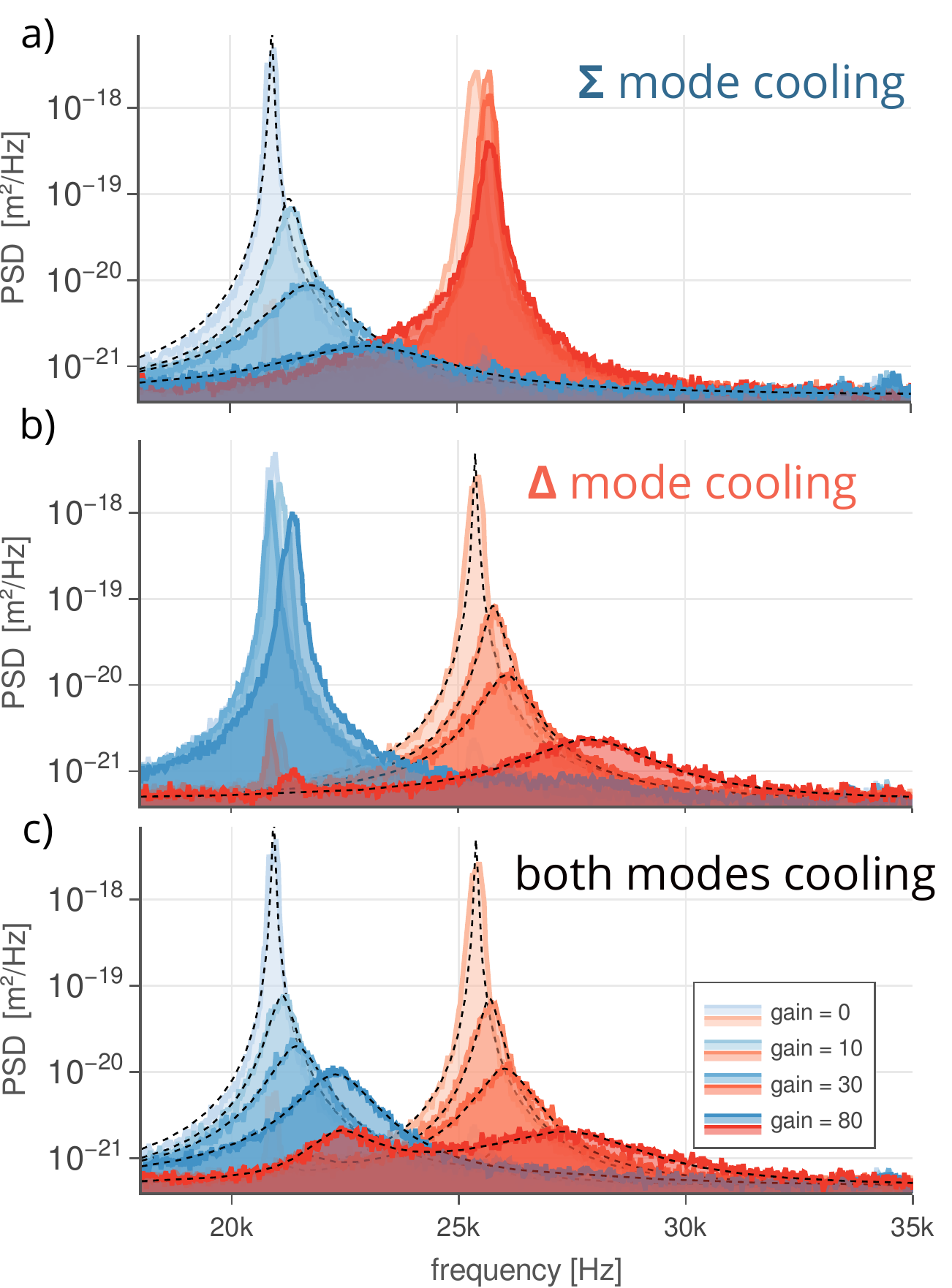}
    \caption{\label{fig2} Experimental cold damping of normal modes. With increasing feedback gain, the PSD broadens in width and simultaneously shrinks in area. a) and b) shows PSDs of both modes when the center of mass ($\Sigma$) and the breathing ($\Delta$) mode, respectively, are cooled independently. c) Both modes are cooled simultaneously. The dashed line denotes the fit of PSD using Eqs. (\ref{psd_c}) and (\ref{psd_b}). The external electric field necessarily couples the normal modes causing the height of the second peak to increase as well. The shift of the resonant oscillatory frequencies is caused by the finite time delay in the feedback loop $\tau = 5.6\,\mu$s. Experimental parameters: ambient pressure $p=0.2$\,mbar, particle charges $q_1 = 88$e and $q_2 = 6$e.}
\end{figure}

\section{Results and discussion}

We applied a controllable, time-dependent electrostatic force to the net charges carried by the optically trapped nano-particles by applying a voltage to a pair of electrodes enclosing the trap [see Fig~\ref{fig1}a)], to implement the cold damping feedback cooling scheme \cite{Tebbenjohanns_PRL_2019}. 
This allowed us to control the effective temperatures of the normal modes independently. A complete and general analysis for feedback cooling of coupled conservative oscillators is provided in the Supplementary Information. 
To start with, we focus on the case where the system is perfectly symmetrical i.e. the traps and particles are identical. 
The equation of motion of our system of the optically bound particles can be expressed in the normal mode basis as 
\begin{equation}
    m \ddot \avec(t) = -\Kmat \avec(t) - \Zmat \dot \avec(t) - \Gmat \dot \avec(t-\tau) + \fvec^L(t),
\end{equation}
where $\avec=(a_\Sigma, a_\Delta)$ are the amplitudes of the centre of mass and breathing modes, with $\dot \avec$ and $\ddot \avec$ the first and second derivatives with respect to time, $\Kmat=\mathrm{Diag}(K_\Sigma, K_\Delta)$ are the modal stiffness, $\Zmat=\mathrm{Diag}(\xi_\Sigma,\xi_\Delta)$ are the effective damping coefficients ($\xi_\Sigma \neq \xi_\Delta$ due to hydrodynamic coupling), $\fvec^L$ is the Langevin force and finally, $-\Gmat \dot \avec(t-\tau)$ is cold damping force, deriving from the external electric force, and $\tau$ is a time delay incurred due to finite, experimental response times. Since this electric force acts on both particles, in proportion to their differing charges, it necessarily couples the normal modes:
\begin{equation}
\fvec^\mathrm{fb}(t) = \begin{bmatrix} f^\mathrm{fb}_\Sigma \\ f^\mathrm{fb}_\Delta \end{bmatrix} = g\begin{bmatrix}
c_\Sigma q_\Sigma & c_\Delta q_\Sigma \\ c_\Sigma q_\Delta & c_\Delta q_\Delta,
\end{bmatrix} \dot \avec (t-\tau) \equiv g' \begin{bmatrix}
    c_\Sigma & c_\Delta \\ c_\Sigma r & c_\Delta r
    \end{bmatrix} \dot \avec (t-\tau) \equiv -\Gmat \dot \avec(t-\tau),  \label{coupling}
\end{equation}
where $g$ is global gain, and $c_\Sigma$, $c_\Delta$ are modal gains, $r$ is the charge ratio, $r=q_\Delta/q_\Sigma$ and $g'=gq_\Sigma$ is the effective scalar gain. 

We measured the delay of our feedback loop to be $\tau = 5.6\,\mu$s which is approximately $1/10$ of the time period of the oscillations in the system.
Our linear feedback signal was implemented on an FPGA card, see Fig.~\ref{fig1}a). 
In the majority of the experiments, we use a low-pass filter whose cut-off frequency was set to above 100\,kHz which is far from the normal mode frequencies, such that we achieve a flat phase response. 
In the case of the lowest pressure ($p= 5\times10^{-3}$ mbar) we applied bandpass filters centred around modal frequencies to get rid of low-frequency noise and detection crosstalks with $y$ axis. 
We calculate instantaneous mode velocities numerically and, for the products, we multiplied modal gains $c_\Sigma$ and $c_\Delta$. 
The final feedback voltage from the FPGA card was amplified and applied to one of the electrodes (the second one was grounded). 

We investigated the cooling performance as a function of the gas pressure and feedback gain to explore the limitations of the method.
In Fig.~\ref{fig2}, we show the single-sided PSDs for both modes (blue colour denotes the CoM i.e. $\Sigma$ mode, red colour denotes the BR i.e. $\Delta$ mode). 
Three different combinations were tested. In particular, we applied the cold damping feedback cooling scheme on each of the modes separately (e.g., $c_\Sigma = 1$ or $c_\Delta = 0$), see Fig~\ref{fig2}a and \ref{fig2}b, or we applied feedback for both modes simultaneously ($c_\Sigma = c_\Delta = 1/\sqrt{2}$), see Fig~\ref{fig2}c. 
These scenarios are described in Supplemental IV B.

The measured signal corresponds to the power spectral densities (PSD), $S^{xx}_{\Sigma/\Delta} $, where the superscript $xx$ indicates the positional PSD, as opposed to the PSD for velocities which we denote, $S^{vv}_{\Sigma/\Delta}$. The PSDs are added to a spectrally flat noise floor, $S_\mathrm{nn}$, associated with noise in our quadrant photodiode detectors and to a further contribution, $S^\mathrm{fb}_{\Sigma/\Delta}$, which quantifies the effective heating of the particles due to measurement noise entering the feedback loop (see Supplemental III). 
We extracted the damping coefficients $\Zmat$ and effective feedback-induced damping rates $\Gmat$ by fitting the PSDs of modes :

\begin{subequations}
\begin{align}
    S^{xx}_{\Sigma}&= \frac{2k_\mathrm{B} T \xi_0}{|P_\Sigma P_\Delta +\Omega^2G_{21}G_{12}e^{-2i\Omega\tau}|^2} (| P_\Delta |^2 + \Omega^2G_{12}^2) + S^\mathrm{fb}_\Sigma (S_\mathrm{nn}) + S_\mathrm{nn}, \label{psd_c}   \\
    S^{xx}_{\Delta} &=  \frac{2k_\mathrm{B} T\xi_0 }{|P_\Sigma P_\Delta +\Omega^2G_{21}G_{12}e^{-2i\Omega\tau}|^2} (|P_\Sigma |^2+\Omega^2G_{21}^2) + S^\mathrm{fb}_\Delta (S_\mathrm{nn}) +  S_\mathrm{nn},  \label{psd_b}
\end{align}
\end{subequations}   
where $\Omega$ is the frequency and $\xi_0 = 6\pi\mu \rho$ is the usual Stokes drag. The contributions from noise-induced heating are,
\begin{subequations}
    \begin{align}
        S^\mathrm{fb}_{\Sigma} &= \Bigg(\frac{| P_\Delta |^2 + 2r \Omega G_{12}\Im(P_\Delta e^{i\Omega \tau}) + r^2 \Omega^2G_{12}^2}{|P_\Sigma P_\Delta + \Omega^2 G_{12} G_{21} e^{-2i\Omega \tau}|^2} \Bigg) \Omega^2g'^2 S_\mathrm{nn}, \label{Snnc}\\
        S^\mathrm{fb}_{\Delta} &= \Bigg(\frac{r^2|P_\Sigma |^2 + 2r\Omega G_{21}\Im(P_\Sigma e^{i\Omega\tau}) + \Omega^2 G_{21}^2}{|P_\Sigma P_\Delta + \Omega^2G_{12} G_{21}e^{-2i\Omega \tau}|^2}\Bigg) \Omega^2g'^2 S_\mathrm{nn}, \label{Snnb}
    \end{align}
\end{subequations}
where $P_{\Sigma/\Omega}$ are quadratic in $\Omega$,
\begin{equation}
    P_\Sigma(\Omega)=m \left[\left(\Omega^2-\frac{G_{11}\sin(\Omega \tau)}{m}\Omega - \frac{K_\Sigma}{m}\right)-\frac{i\Omega}{m}\big(\xi_\Sigma+G_{11}\cos(\Omega\tau) \big)  \right],\label{psd_cb}
\end{equation}
\begin{equation}
    P_\Delta(\Omega)=m \left[\left(\Omega^2-\frac{G_{22}\sin(\Omega \tau)}{m}\Omega - \frac{K_\Delta}{m}\right)-\frac{i\Omega}{m}\big(\xi_\Delta+G_{22}\cos(\Omega\tau) \big)  \right]. \label{psd_cb2}
\end{equation}

The resonant peaks appearing in the PSD correspond approximately to the minima of $|P_\Sigma(\Omega)|^2$ and $|P_\Delta(\Omega)|^2$. When $\tau=0$ this gives peaks at $\Omega_\Sigma\approx\sqrt{K_\Sigma/m}$ and $\Omega_\Delta\approx\sqrt{K_\Delta/m}$. However, when the time delta, $\tau$, is appreciable, the resonant frequencies shift significantly with increasing gain, see Fig.~\ref{fig2} and the Supplemental IV C.

\begin{figure*}[htb] 
    \centering
    \includegraphics[width = \textwidth]{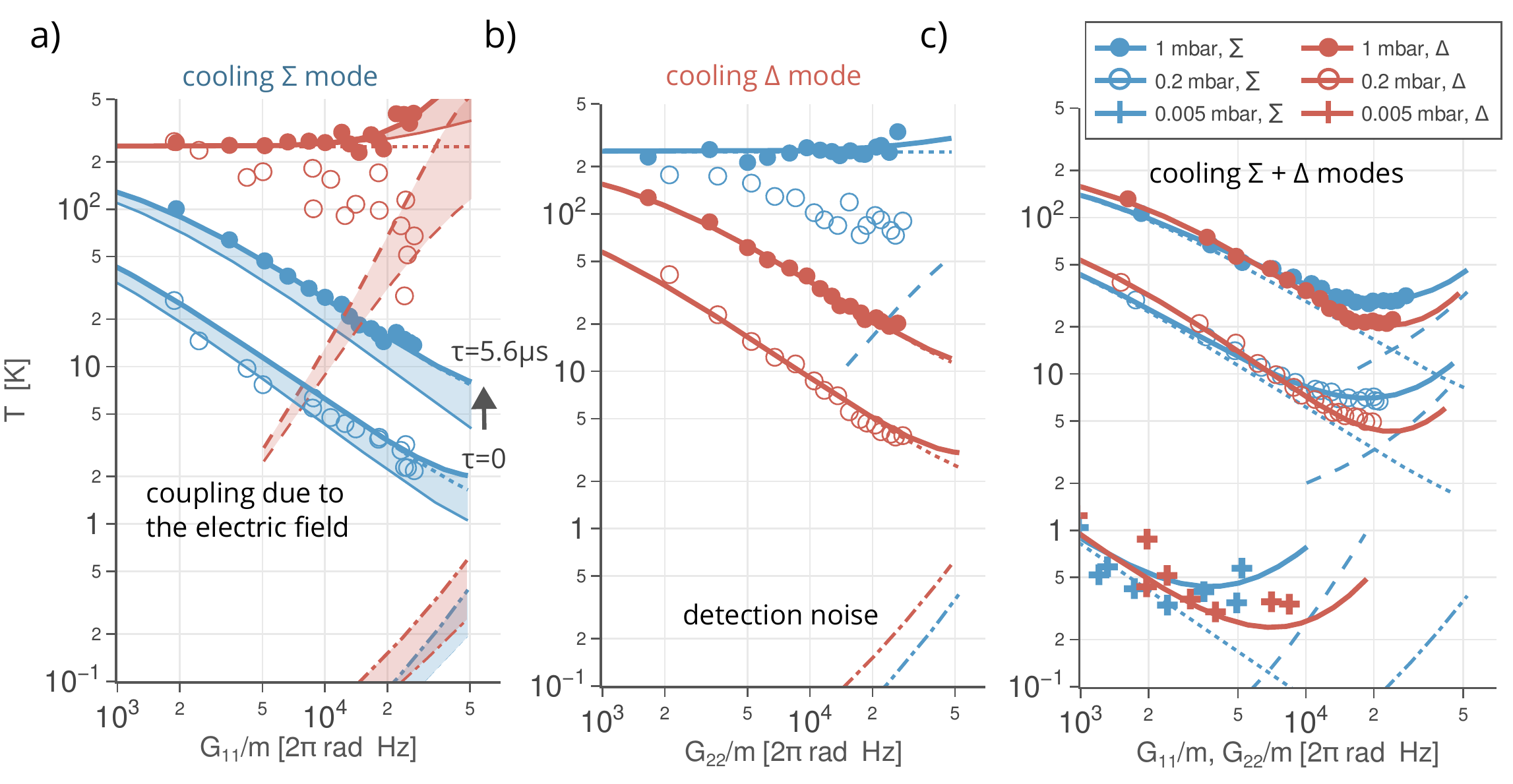}
    \caption{\label{fig3} Cooling performance visualized using the effective temperatures of the modes. The values of the feedback damping $\Gmat$ coefficients were determined by fitting the experimental PSDs,  $S^{xx}_{\Sigma/\Delta}$.  a) and b) shows the effective temperatures when the centre of mass and the breathing mode, respectively, are cooled independently. c) Both modes cooled simultaneously. Sub-Kelvin effective temperature was achieved for both modes at $5 \times 10^{-3}$\,mbar.  Full curves show temperatures determined from theoretical PSDs [Eqs. (\ref{psd_c}) and (\ref{psd_b})], dotted curves show the temperature when the values of coupling terms were set to be zero $G_{12} = G_{21} = 0$, dashed and dash-dotted curves illustrate heating of the motion of the particles due to the coupling induced by the external field and due to the measurement noise fed back to control feedback loop, respectively. The filled areas in part a) illustrate the increase of the effective temperatures caused by feedback loop delay $\tau = 5.6\,\mu$s, i.e. the lower values of temperatures are calculated for zero delay. (for $p = 1$ and 0.2 mbar, $q_1 = 88$e and $q_2 = 6$e, for $p = 5 \times 10^{-3}$\,mbar, $q_1 = 27$e and $q_2 = 1$e). }
\end{figure*}

As expected, as we increase the value of the global feedback gain $g$, the PSD for cooled mode broadens in width and simultaneously shrinks in area, see Figs~\ref{fig2}. 
When the global gain, $g$, is zero, each modal PSDs contain a single peak, corresponding to the resonant frequency of the associated mode. However, as $g$ is increased, a small, subsidiary peak begins to appear at the resonant frequency of the second mode, indicating a degree of feed-back induced inter-modal coupling,  see Figs.~\ref{fig2}a) and \ref{fig2}b).  
This limits the effect of the feedback cooling scheme. 
For independently cooled modes, this effect would be absent in a perfectly symmetric system (with identical particles and traps), see Supplemental information. 
In contrast, when cooling both modes simultaneously, the inter-modal coupling is introduced via off-diagonal terms of the gain matrix, $\Gmat$, and is completely unavoidable here, see Fig.~\ref{fig2}c).

To quantify the cooling performance we calculated the effective temperatures $T_{\Sigma/\Delta}$ of the oscillation modes from the measured displacement time traces $a_\Sigma(t)$ and  $a_\Delta(t)$. 
To do this we used the kinetic energy of the oscillators 
\begin{equation}
    T_{\Sigma/\Delta} = 2\langle E^\mathrm{kin}_{\Sigma/\Delta} \rangle /k_\mathrm{B} = m \langle \dot{a}_{\Sigma/\Delta}^2 \rangle /k_\mathrm{B}, 
\end{equation}    
for which the equipartition theorem is valid even for anharmonic trapping potentials \cite{Hebestreit_RSI_2018}.
For obtained $\langle \dot{a}_{\Sigma/\Delta}^2 \rangle$  we numerically integrate the power spectral density for mode velocity which can be expressed using power spectral density for particle displacement,
\begin{equation}
    \langle \dot{a}_{\Sigma/\Delta}^2 \rangle = \int_0^\infty S^{vv}_{\Sigma/\Delta}(\Omega) \mathrm{d}\Omega = \int_0^\infty \Omega^2 [S^{xx}_{\Sigma/\Delta}(\Omega) - S_\mathrm{nn}] \mathrm{d}\Omega.
\end{equation}
It should be noted that we have subtracted the detection noise floor $S_\mathrm{nn}$ from our experimentally determined power spectral density.

Figures~\ref{fig3}a) -- \ref{fig3}c) show the magnitude of the effective temperature for different values of gas pressure ($ p = 1,\, 0.2$ and $5 \, \times 10^{-3}$\, mbar) determined for increasing values of feedback induced damping $G_{11} = g c_\Sigma q_\Sigma$ and $G_{22} = g c_\Delta q_\Delta$ which was controlled, during the experiment, by the magnitude of the global feedback gain $g$. 
Complementary to the power spectral densities presented in  Fig.~\ref{fig2}, we compare here three cases, i.e. independent cooling of the modes [$c_\Delta = 0$ in Fig.~\ref{fig3}a)  and $c_\Sigma = 0$ in Fig.~\ref{fig3}b)] and simultaneous cooling of both modes Fig.~\ref{fig3}c). At the lowest studied pressure ($5 \, \times 10^{-3}$\, mbar) we achieved sub-Kelvin temperature for both modes.  
For all the studied cases, there is an optimal value of feedback damping rate $G_{ii}/m$ for which we reach minimal values of effective temperature for a particular surrounding gas pressure.

\begin{figure*}[htb]
    \centering
    \includegraphics[width = \textwidth]{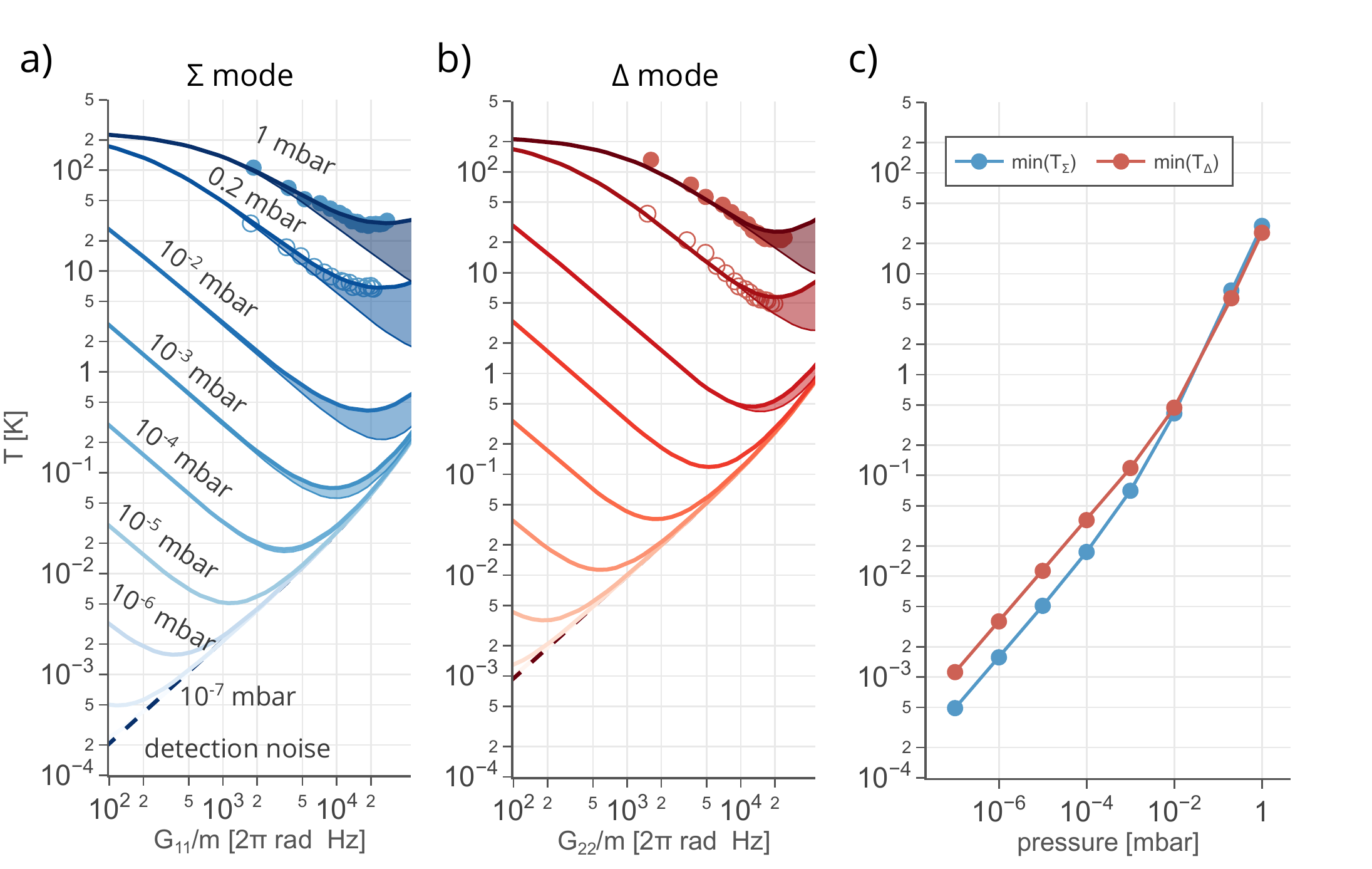}
    \caption{\label{fig4} Extrapolated cooling performance visualized using the effective temperatures of CoM mode a) and BR mode b). The dashed line illustrates heating due to the noise in detection, which differs for both modes see Eqs.~(\ref{Snnc}) and (\ref{Snnb}). The coloured area shows the effect of coupling between modes due to the electric feedback force. c) The minimal values of effective temperatures as a function of the gas pressure determined from a) and b).}
\end{figure*}

For independently cooled modes [Figs.~\ref{fig3}a) and \ref{fig3}b)] the presence of an optimal condition arises due to competition between two mechanisms: the cooling effect to increasing feedback damping, and the inevitable heating effect caused by feeding back measurement noise into the control loop, $T^\mathrm{det}_{\Sigma/\Delta} \propto \Gmat S_\mathrm{nn}$ (see dash-dotted lines).  In this respect, the limits in the efficiency of independently cooling the normal modes are similar to those operating in the single particle case \cite{Tebbenjohanns_PRL_2019}.

When independently cooling the CoM mode ($c_{\Sigma} \neq  0$ and $c_{\Delta} = 0$), the electric force, Eq.~(\ref{coupling}), acting on the CoM mode is not influenced by the breathing mode, $f_\Sigma = g [c_{\Sigma} q_{\Sigma} v_{\Sigma} + c_{\Delta} q_{\Sigma} v_{\Delta}] = g c_{\Sigma} q_{\Sigma} v_{\Sigma}$, on the other hand, there is a coupling between un-cooled breathing mode and CoM mode, mediated by feedback electric force $f_\Delta = g [c_{\Sigma} q_{\Delta} v_{\Sigma} + c_{\Delta} q_{\Delta} v_{\Delta}] = g c_{\Sigma} q_{\Delta} v_{\Sigma}$. 
This resulting variance in the BR mode is,
\begin{equation}\label{eq:varb}
\langle a^2_\Delta \rangle = \frac{k_\mathrm{B}T}{K_\Delta}+2k_\mathrm{B}T\xi_\Delta g^2 q_\Delta^2 \int^\infty_\infty d\Omega \frac{\Omega^2}{|P_\Sigma|^2|P_\Delta|^2}. 
\end{equation}
The first term in Eq. (\ref{eq:varb}) is constant, while the second term is necessarily positive and dependent on the gain. This second term is connected with the heating of the uncooled mode, due to feedback-induced coupling.
Indeed, the heating of the un-cooled breathing mode was observed experimentally for 1\,mbar, see the red line for the total value of total effective temperature and the red dashed line which illustrates the heating due to the coupling of modes by the electric field, in Fig.~\ref{fig3}a). 
However, for lower pressure, we observed an opposite trend, which we observed also in our numerical model by assuming very subtle asymmetry in our system, where our idealized modes are not fully valid, see Supplemental information for a full description. 
The asymmetry in our numerical model was induced,  by considering only a very small difference in particle sizes $<$ 0.5\%, which is much smaller than the standard deviation of particle size specified by a manufacturer (3\%). 
The non-zero value of time delay $\tau$ in our feedback loop caused the increase of values of the effective temperatures $T$ (about $2\times$), as illustrated using colour-filled areas in Fig.~\ref{fig3}a).

Figure~\ref{fig3}c) compares the values of the effective temperatures when both modes were cooled simultaneously. For the sake of clarity, the heating mechanisms are here illustrated only for the CoM mode. 
Since the heating due to coupling between the modes, scales with damping of the surrounding gas [e.g., for CoM it is  $ 2k_BT\xi_0  \Omega^2G_{12}^2$, see Eq.~(\ref{psd_c})], it is the main limiting mechanism at pressures higher than $10^{-3}$\,mbar, on the other hand  the heating due to measurement noise which is proportional to $\Omega^2g'^2 S_\mathrm{nn} $ [see Eqs.~(\ref{Snnc}) and (\ref{Snnb})] starts to be important mainly at the lower pressure, see also Figs.~\ref{fig4}a) and  \ref{fig4}b) for values of effective temperature for both simultaneously cooled modes extrapolated for low pressure [other parameters are same as in Fig.~\ref{fig3}c)]. 
Note, that the heating due to the detection noise differs for both modes, see Eqs.~(\ref{Snnc}) and (\ref{Snnb}), and thus the minimal values of effective temperatures determined from curves presented in Figs.~\ref{fig4}a) and  \ref{fig4}b) differs too, see Fig~\ref{fig4}c) for pressure dependence of minimal values of effective temperatures. 

Finally, a numerical parametric survey of optimal cooling conditions is provided in Supplemental IV B. 
The main conclusions are as follows. First, for a fixed total charge, $|q_1|+|q_2|$, cooling is most efficient when all the charge is located on one particle. This is true whether we are cooling the modes separately or together. In the former case, the cooling mechanism is purely sympathetic (c.f. \cite{Arita_Optica_2022}), with the uncharged particle cooled indirectly, via optical binding forces. 
Second, the CoM mode is most efficiently cooled with $c_\Sigma=1$ and $c_\Delta=0$ while BR mode is most effectively cooled when $c_\Sigma=0$ and $c_\Delta=1$. Third, for symmetric systems with $q_1=q_2$ the breathing mode cannot be cooled in the manner discussed here. Similarly, when $q_1=-q_2$ the CoM mode cannot be cooled.

\section{Conclusion}

In summary, we have demonstrated cooling of the motion of optically interacting levitated nanoparticles to sub-Kelvin effective temperatures using a cold damping method employing an external electric field. 
We have discussed the limitations of the cooling method using a rigorous theoretical model. 
For individually cooled motional modes of coupled oscillators, the most stringent limit is due to the detection efficiency and its noise, which is analogous to the case of single levitated particles. 
In the case of simultaneous cooling of the modes, the coupling between the modes by the feedback electric force is an additional source of feedback-induced heating which limits the performance of the method at pressures higher than $10^{-3}$\,mbar.  At lower pressures, the detection noise is the dominant heating mechanism.
Thus, in principle, the cold damping method could get closer to the quantum ground state with an optimized detection efficiency, operating at lower pressures.
Motional cooling of several interacting levitated nanoparticles is a significant step towards the generation of cooled arrays of optically levitated particles, which will allow multi-particle studies at the boundary of classical and quantum physics.

\bibliography{tweeznew}

\end{document}